\documentclass[10pt, conference]{IEEEtran}
\usepackage[a4paper,top=2cm,bottom=3.5cm,left=1.6cm,right=1.6cm]{geometry}

\usepackage{makecell}
\usepackage{hyperref}
\usepackage{array}
\usepackage{graphicx,amssymb,amsmath}
\usepackage{multicol}
\usepackage[noadjust]{cite}
\usepackage{setspace}
\usepackage{subfigure}
\usepackage{float}
\usepackage{url}
\usepackage{tikz}
\usetikzlibrary{arrows.meta, positioning, shapes.multipart}
\usepackage{stfloats}
\usepackage{amsthm,pifont}
\usepackage{flushend}
\usepackage{cases,subeqnarray}
\usepackage{bm,multirow,bigstrut}
\usepackage{amsmath, amsthm, amssymb}
\usepackage{textcomp}
\usepackage{latexsym,bm}
\usepackage{booktabs}
\usepackage{xcolor}
\usepackage{mathtools}
\usepackage{dsfont}
\usepackage{extarrows}
\usepackage{epsfig}
\usepackage{epsfig}
\usepackage{epstopdf}
\usepackage{colortbl}
\usepackage{algpseudocode}
\usepackage{tcolorbox}
\usepackage{algorithmicx,algorithm}

\theoremstyle{plain}

\theoremstyle{plain}

\usepackage{amsmath}

\definecolor{Gray}{gray}{0.85}

\def\BibTeX{{\rm B\kern-.05em{\sc i\kern-.025em b}\kern-.08em
    T\kern-.1667em\lower.7ex\hbox{E}\kern-.125emX}}

\begin{document}

\title{Model Context Protocol-based Internet of Experts
For Wireless Environment-aware LLM Agents}
\author{Zongxi Liu\textsuperscript{1}, Hongyang Du\textsuperscript{2}
\\
\small{\textsuperscript{1}School of Electronic Science and Engineering, Nanjing University, Nanjing, China}\\
\small{\textsuperscript{2}Department of Electrical and Electronic Engineering, University of Hong Kong, Hong Kong SAR, China}\\
\small{Emails: 522022230067@smail.nju.edu.cn, duhy@eee.hku.hk}
}
\vspace{-2cm}
\maketitle
\vspace{-2cm}
\begin{abstract}
Large Language Models (LLMs) exhibit strong general-purpose reasoning abilities but lack access to wireless environment information due to the absence of native sensory input and domain-specific priors. Previous attempts to apply LLMs in wireless systems either depend on retraining with network-specific data, which compromises language generalization, or rely on manually scripted interfaces, which hinder scalability. To overcome these limitations, we propose a Model Context Protocol (MCP)-based Internet of Experts (IoX) framework that equips LLMs with wireless environment-aware reasoning capabilities. The framework incorporates a set of lightweight expert models, each trained to solve a specific deterministic task in wireless communications, such as detecting a specific wireless attribute, e.g., line-of-sight propagation, Doppler effects, or fading conditions. Through MCP, the LLM can selectively query and interpret expert outputs at inference time, without modifying its own parameters. This architecture enables modular, extensible, and interpretable reasoning over wireless contexts. Evaluated across multiple mainstream LLMs, the proposed wireless environment-aware LLM agents achieve $40\%$-$50\%$ improvements in classification tasks over LLM-only baselines. More broadly, the MCP-based design offers a viable paradigm for future LLMs to inherit structured wireless network management capabilities.
\end{abstract}
\IEEEpeerreviewmaketitle
\section{Introduction}
Large Language Models (LLMs) have progressed from the 117‑million‑parameter GPT-1 to trillion‑scale, multimodal systems such as GPT-4, Llama‑3.1, and Gemini. Their scaling unlocks reliable chain-of-thought reasoning, code synthesis, and long-horizon planning, enabling agentic platforms that automate software development, legal draft analysis, medical triage, and industrial inspections~\cite{zhou2024large,bi2024deepseek}.
These reasoning abilities extend to wireless networks, allowing LLM agents to convert network service goals into scheduling and beamforming commands, forecast congestion for prompt load reallocation, and interpret unusual fading signatures for maintenance~\cite{zhou2024large}. An agent-driven control plane built on these functions reacts faster than rule-based scripts and learns directly from diverse telemetry, without the need for handcrafted features~\cite{wang2024survey}.

\begin{figure}[t]
\centering
\includegraphics[width = 0.45\textwidth]{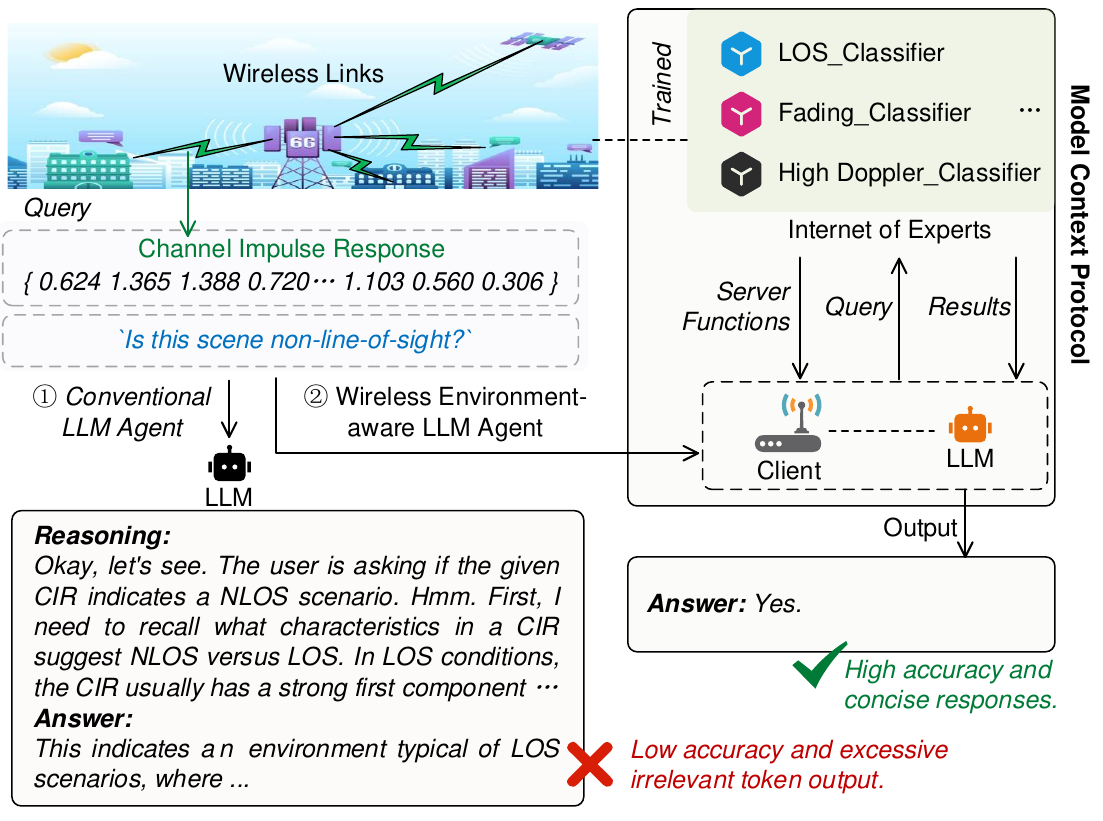}
\caption{Illustration of LoS or NLoS environment classification query where a conventional LLM agent generates low-accuracy and verbose responses, compared to an MCP-based internet of experts for wireless environment-aware LLM agents that achieve high accuracy and concise results.}
\label{fig_mo}
\end{figure}
However, the core working paradigm of LLMs, i.e., predicting the next token in a natural-language corpus, offers no inductive bias for complex-valued baseband signals or logarithmic metrics that dominate radio engineering~\cite{zhou2024large}. For example, as shown in the left-hand side of Fig.~\ref{fig_mo}, when tasked with predicting quantities such as path loss or Doppler shift, a vanilla model may output physically impossible values, mishandle unit operations like the linear addition of decibels, or overlook critical phase information, ultimately undermining reliability in automated control systems.
Efforts to compensate generally follow two primary paths: {\textit{1) Prompt-centric schemes}} wrap telemetry into textual templates and rely on chain-of-thought reasoning to derive actions~\cite{zhou2025large}. However, the generation remains fundamentally uncertain, with no guarantee that the output adheres to physical laws, leaving LLMs vulnerable to hallucination even when the prompts are carefully crafted.
{\textit{2) Domain-oriented fine-tuning}} trains LLMs on annotated traces and protocol documents~\cite{liu2025llm4wm}, which improves performance in familiar scenarios. Yet, this approach faces significant barriers: collecting high-quality domain data is expensive and time-consuming, and fine-tuning requires billions of gradient updates, resulting in substantial energy consumption. More importantly, the approach lacks scalability. Whenever deployment conditions change, such as carrier frequency shifts, new antenna arrays, or emerging mobility patterns, the LLM must undergo re-fine-tuning on new datasets, making it impractical for dynamic and evolving wireless environments~\cite{noh2025adaptive}.
As a result, neither approach enables real-time adaptation to key wireless dynamics such as Line-of-Sight (LoS) transitions, fast fading, or bursty interference, leaving LLM-based reasoning detached from the environment it is intended to manage~\cite{du2024reinforcement}.

To bridge this gap, an LLM agent must pair its linguistic reasoning with verifiable wireless perception: for any user query, the model should identify the relevant channel features, obtain them through specialized analytics, and integrate the results into its response. The recently released Model Context Protocol (MCP) addresses this need by exposing a uniform JSON-RPC interface that lets language models discover, call, and compose external tools hosted on independent {\textit{``servers''}}~\cite{hou2025model,anthropic_mcp_2025}. Public MCP hubs already expose generic utilities, e.g., reverse geocoding, document retrieval, code execution sandboxes, and calculator endpoints, through machine-readable schemas that return compact outputs compatible with an LLM context window. Building on this infrastructure, we introduce a wireless‑oriented Internet of Experts (IoX) in which each MCP server hosts a deterministic wireless analyser, such as a path‑loss estimator, a deep‑reinforcement‑learning power‑allocation policy, or a beam‑search throughput predictor, as shown on the right-hand side of Fig.~\ref{fig_mo}. 
Because the experts run on edge or cloud nodes and communicate only low-dimensional tensors, an LLM agent can issue a perception or control query, receive the answer with a low latency, and fuse it into its next token stream without retraining its frozen weights~\cite{barati2016initial}. This arrangement pairs the linguistic flexibility and safety alignment of the foundation model with real-time channel intelligence, closing the perception–reasoning gap that limits conventional autonomous wireless agents.

Building on the above motivation, we introduce a complete system design that integrates LLM agents with wireless environment perception using the MCP and a pool of expert tools. Our contributions are:
\begin{itemize}
\item We formulate the IoX as a modular architecture where each wireless attribute is associated with a lightweight, task-specific expert model. These experts are trained independently to detect scene features such as LoS conditions, Doppler shifts, or user mobility.
\item We develop an MCP-based framework in which the LLM autonomously selects, queries, and interprets the relevant experts during task execution. This enables the agent to ground its high-level decisions in accurate, runtime scene observations without requiring retraining or prior embedding of environment-specific knowledge.
\item We implement and evaluate a proof-of-concept system that demonstrates the effectiveness of IoX in a wireless setting. Experimental results show that MCP-based querying of expert classifiers supports accurate scene identification, enabling the LLM agent to generate responses aligned with wireless environmental conditions.
\end{itemize}

\section{System Model}
In this section, we introduce the overall system architecture and present a formal problem formulation that guides our design. We first describe how the LLM agent interacts with external wireless expert tools through the MCP. We then formulate the agent’s goal as a structured decision problem that integrates both natural-language prompts and real-time wireless scene information.

\subsection{System Overview}
The system consists of three components:
\begin{enumerate}
\item An LLM agent that interprets user queries and generates task-specific responses;
\item An MCP interface that enables the LLM to call these tools during inference, which is explained in detail in Section~\ref{mcpde};
\item A set of wireless network management experts, which may include AI models or rule-based analyzers~\cite{zappone2019model}. Without loss of generality, here we consider the expert pool to contain $M$ independent classifiers $\{\mathcal{E}_m\}_{m=1}^{M}$, where each $\mathcal{E}_m$ estimates the probability of a specific scene attribute $s_m$ being present. These attributes include typical propagation and mobility conditions of wireless environments, such as LoS, high Doppler shift, or small-scale fading of wireless signals.
\end{enumerate}
The input to the system is a user query that requests a deterministic result from the LLM agent, i.e., $q\in\mathcal{Q}$, and a current wireless observation, typically represented as a complex channel impulse response $\mathbf{h} \in \mathbb{C}^N$. 

\subsection{Problem Formulation}
Let $r_{\pi}\left(q,\mathbf{h}\right)$ denote the response of an LLM agent.  
For each input, i.e., $\left(q,\mathbf{h}\right)$, a ground-truth reply $y\left(q,\mathbf{h}\right)$ could be obtained from measurement or simulation. We define the accuracy metric as
\begin{equation}
A\left(r_{\pi},y\right)=
\begin{cases}
1, & r_{\pi} = y,\\
0, & r_{\pi} \neq y,
\end{cases}
\end{equation}
where $\pi$ is the agent policy and $\mathcal{D}$ is the joint distribution of queries and channel states.
In typical settings, the policy $\pi$ is to directly map the input pair to a response using a standalone LLM. We aim to design an MCP-based policy in which the LLM augments its reasoning by querying a set of wireless expert tools, enabling environment-aware decision-making for
\begin{align}
\max_{\pi}\;
&{\mathbb{E}}_{(q,\mathbf{h})\sim\mathcal{D}}\left[A\!\left(r_{\pi}\left(q,\mathbf{h}\right),\,y\left(q,\mathbf{h}\right)\right)\right].
\label{eq:objective}
\end{align}
Specifically, when reasoning over a prompt, the LLM selectively issues queries to relevant experts and receives structured outputs for response generation. Experts can be deployed on edge or cloud servers, and their low-dimensional outputs ensure that inference remains efficient.


\section{MCP-based Internet of Experts}
In this section, we introduce the training process of the IoX and the design of the wireless environment-aware LLM agent based on the MCP. The IoX framework aims to build a collection of specialized expert models by categorizing wireless scenarios and training targeted lightweight networks. MCP serves as the coordination protocol that enables the LLM agent to query, select, and integrate expert outputs for context-aware reasoning and decision-making. 

\subsection{IoX Training}\label{subsec:iox_training}

\begin{figure*}[t]
\centering
\includegraphics[width = 0.9\textwidth]{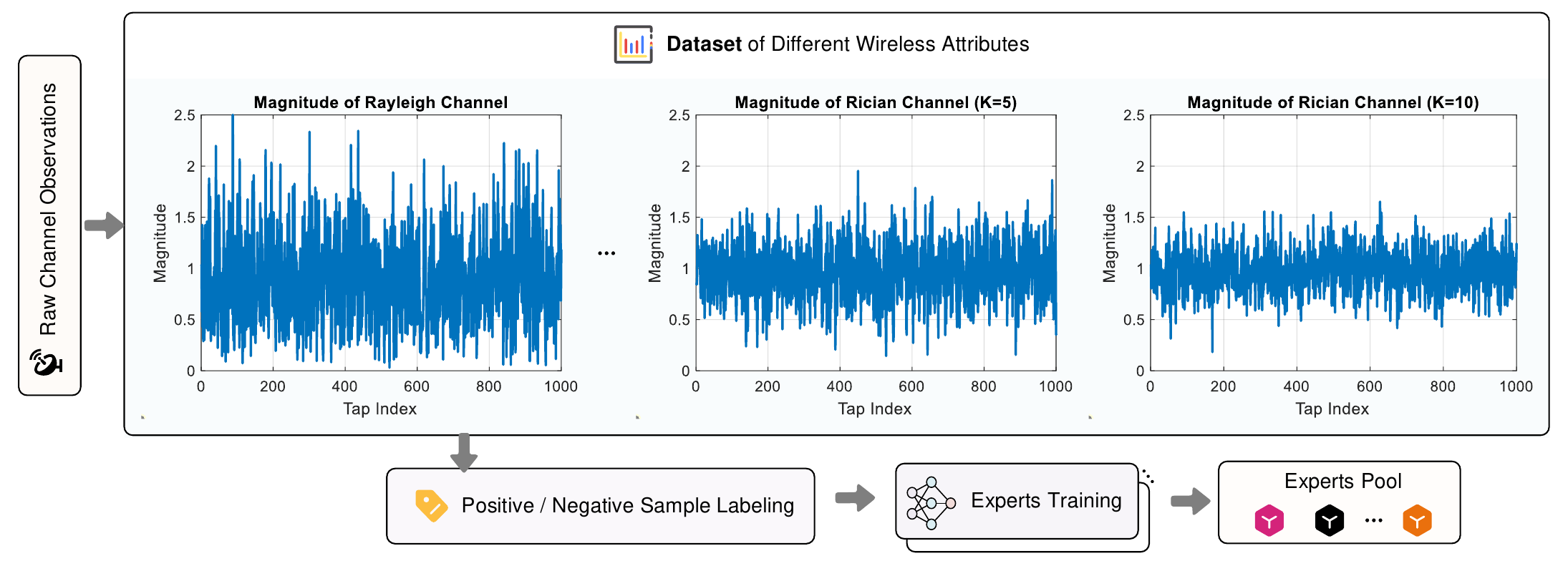}
\caption{The training process of IoX. Raw channel observations are organized into wireless attribute-specific datasets, where positive and negative samples are labeled based on scene conditions. Each expert is trained independently using a lightweight MLP and then integrated into a modular expert pool for LLM reasoning via MCP.}
\label{fig_iox}
\end{figure*}

The IoX is designed to provide the LLM with structured, interpretable wireless context labels while remaining deployable on edge devices with limited computation. Each attribute \( s_m \in \mathcal{S} \), such as a specific Doppler shift level, a fading profile, e.g., Rayleigh or Rician with a given \( K \)-factor, or a propagation type, e.g., LoS or non-LoS, is treated as a standalone classification problem~\cite{panic2013fading}. This decomposition ensures that each expert specializes in detecting a clearly defined physical phenomenon, enabling modular training and independent evaluation.

\textbf{1) Scene-centric data construction.}
To isolate the detection of each wireless condition, we construct attribute-specific datasets \( \mathcal{D}_m = \{(\mathbf{h}_i, y_i)\} \), where \( \mathbf{h}_i \in \mathbb{R}^n \) represents the real-valued magnitude vector derived from a complex-valued channel impulse response and \( y_i \in \{0,1\} \) denotes whether the scene corresponding to the attribute \( s_m \) is present. Positive samples are drawn from a synthetic or measured dataset corresponding to that attribute, e.g., Rician fading \( K=10 \), while negative samples are sampled uniformly from all other non-matching scenes. This strategy turns the overall multi-label recognition problem into a set of well-conditioned, balanced binary tasks. It avoids label imbalance issues and simplifies training dynamics, while enabling flexible expert pool extension, i.e., introducing a new scene class only requires additional positive examples and resampling of negatives, with no need to retrain the rest of the system.

\textbf{2) Lightweight expert backbone.}
Each expert \( \mathcal{E}_m \) is implemented as a compact Multi-Layer Perceptron (MLP) to minimize memory and latency overhead as
\begin{equation}
f_{\theta_m}\!(\mathbf{h})\!=\!\sigma\!\left( W_3 \!\cdot\! \text{ReLU}(W_2 \!\cdot\! \text{ReLU}(W_1 \!\cdot\! \mathbf{h}\! + \!b_1)\! +\! b_2)\! + \!b_3 \right)\!,
\end{equation}
where $\sigma$ is the sigmoid activation, $W_i$ and $b_i$ are weight matrices and bias vectors of the $i_{\rm th}$ layer, and $\mathbf{h}$ is the channel magnitude vector input. The architecture uses two ReLU layers and a final sigmoid output to express non-linear decision boundaries while keeping the number of parameters small. The experiment confirms that dozens of experts can be evaluated per frame on a standard mobile GPU, allowing the LLM to invoke multiple expert queries simultaneously without violating real-time constraints.

\textbf{3) Independent optimization and continual extension.}
Each expert is trained using a cross-entropy loss as
\begin{equation}
\mathcal{L}_m = - \frac{1}{N} \sum_{i=1}^{N} \left[ y_i \log p_i + (1 - y_i) \log (1 - p_i) \right],
\end{equation}
where $p_i = f_{\theta_m}(\mathbf{h}_i)$. This loss function is standard for binary classification tasks where output probabilities \( p_i \in (0,1) \) are interpreted as confidence scores~\cite{ruby2020binary}. It penalizes both false positives and false negatives in proportion to their deviation from the true label, providing a smooth optimization surface. The use of binary loss instead of soft multi-label objectives avoids inter-task interference and allows each model to be trained, evaluated, and maintained independently~\cite{ruby2020binary}. 

This training setup supports continual extension. When new wireless environment conditions are encountered, new experts can be trained and appended without modifying the LLM or any existing expert. This design avoids catastrophic forgetting and preserves the generalization capability of LLMs, which interact with experts through semantic queries.

The final system consists of a modular expert pool $\{\mathcal{E}_m\}_{m=1}^{M}$, each outputting a posterior probability \( p(s_m \mid \mathbf{h}) \). At inference time, the LLM dynamically selects and queries the relevant experts based on its reasoning chain and uses their outputs to interpret wireless channel conditions. For instance, it may infer {\textit{``non-LoS with high Doppler and moderate $K$-factor Rician fading''}} by combining expert outputs. This decouples physical signal interpretation from high-level reasoning, supports real-time deployment in changing environments, and maintains interpretability by associating each expert decision with a well-defined wireless concept.

\subsection{Model Context Protocol Design}\label{mcpde}
The MCP enables structured interaction between an LLM and external expert tools during inference~\cite{hou2025model}.  
In practice, the LLM runs inside an {\textit MCP host}, which embeds a lightweight {\textit client} library.  
The client translates model-generated JSON into JSON-RPC 2.0 calls, forwards them to remote servers, and streams the replies back into the context~\cite{anthropic_mcp_2025}.  
This arrangement supports real-time reasoning over physical-layer observations without retraining or embedding signal processing code into the model~\cite{anthropic_mcp_2025}.  
In addition to tool invocation, MCP exposes resource and template endpoints, but in this work, we focus on the tool layer that serves wireless perception.

Formally, the LLM is expected to return a decision $r_\pi(q,\mathbf{h})$ that matches the true label $y(q,\mathbf{h})$ by optionally consulting a set of registered experts $\{\mathcal{E}_m\}_{m=1}^{M}$, where each $\mathcal{E}_m$ maps a channel feature vector to a confidence score as
\begin{equation}
\mathcal{E}_m(\mathbf{h}) = p(s_m \mid \mathbf{h}) \in [0,1].
\end{equation}

\begin{figure}[t]
\centering
\begin{tcolorbox}[colback=white!99!gray!1, colframe=black!75, title=MCP-based LLM Agent Host-Client Pipeline]
{\small
\begin{algorithmic}[1]
\Statex \textbf{Step 1: Expert Registration (Offline)}
\State Refer to Fig.~\ref{fig:expert-registration} for registration structure.

\Statex \textbf{Step 2: Expert Planning (Host Planner)}
\State Input: user query $q$ and wireless observation $\mathbf{h}$
\State Planner prompts LLM: ``Which experts are relevant?''
\State LLM returns: \texttt{"mcp\_calls": [$s_1$, $s_3$, ...]}

\Statex \textbf{Step 3: Expert Invocation (MCP Client)}
\State Refer to Fig.~\ref{fig:expert-invocation} for query and response schema.
\For{each expert $s_m$ in \texttt{mcp\_calls}}
\State Construct and send JSON query with input $\mathbf{h}$
\State Await response from expert server
\EndFor

\Statex \textbf{Step 4: Response Handling (MCP Client)}
\For{each expert $s_m$ in \texttt{mcp\_calls}}
\State Receive response:
\State \quad \{``confidence'': $p(s_m \mid \mathbf{h})$, ``status'': OK, ``source\_id'': $m$ \}
\State Store to \texttt{expert\_results}
\EndFor

\Statex \textbf{Step 5: Context Augmentation (Host)}
\State Assemble prompt: query $q$ + \texttt{expert\_results}
\State Inject into LLM context window

\Statex \textbf{Step 6: Final Reasoning (LLM Core)}
\State LLM consumes structured prompt and generates output $r_\pi(q, \mathbf{h})$
\end{algorithmic}}
\end{tcolorbox}
\caption{Runtime pipeline of an MCP-based wireless environment-aware LLM agent. The client logic resides inside the host process and mediates all JSON-RPC exchanges with expert servers, enabling modular and stateless reasoning over wireless signals through expert composition.}
\label{fig:mcp-endtoend}
\end{figure}

To support such coordination, MCP defines a modular and stateless runtime structure inspired by modern tool-invocation APIs~\cite{qintoolllm}. The full pipeline comprises several stages, illustrated in Fig.~\ref{fig:mcp-endtoend}, and described as
\begin{figure}[t]
\centering
\begin{tcolorbox}[colback=white!99!gray!1, colframe=black!75, title=Expert Registration]
{\small \begin{algorithmic}[1]
\State \textbf{function} \textsc{RegisterExpert}($s_m$, description, input\_schema)
\State \hspace{0.3cm} \textbf{define} expert\_entry $\gets$ \{
\State \hspace{0.7cm} \texttt{"name"}: $s_m$,
\State \hspace{0.7cm} \texttt{"description"}: description,
\State \hspace{0.7cm} \texttt{"input\_schema"}: input\_schema
\State \hspace{0.3cm} \}
\State \hspace{0.3cm} Add expert\_entry to MCP registry
\State \textbf{end function}
\Statex \hrulefill
\Statex \textbf{// Example: Registering LoS classifier}
\State $s_m \gets$ \texttt{"detect\_los"}
\State description $\gets$ \texttt{"Returns the probability that the scene is under LoS condition given channel features."}
\State input\_schema $\gets$
\State \hspace{0.3cm} \{
\State \hspace{0.6cm} \texttt{"type"}: \texttt{"object"},
\State \hspace{0.6cm} \texttt{"properties"}: \{
\State \hspace{1.0cm} \texttt{"h"}: \{
\State \hspace{1.3cm} \texttt{"type"}: \texttt{"array"},
\State \hspace{1.3cm} \texttt{"items"}: \{\texttt{"type"}: \texttt{"number"}\},
\State \hspace{1.3cm} \texttt{"description"}: \texttt{"Channel vector of $n$ real values"}
\State \hspace{1.0cm} \}
\State \hspace{0.6cm} \},
\State \hspace{0.6cm} \texttt{"required"}: [\texttt{"h"}]
\State \hspace{0.3cm} \}
\State \textsc{RegisterExpert}($s_m$, description, input\_schema)
\end{algorithmic}}
\end{tcolorbox}
\caption{Expert registration schema in the Internet of Experts. Each expert is associated with a unique identifier, a semantic description, and a JSON-style input specification.}
\label{fig:expert-registration}
\end{figure}

\begin{enumerate}
\item \textbf{Expert registration (offline)}: Each expert is registered with a unique identifier $s_m$, a semantic description, and a JSON-style input schema, as shown in Fig.~\ref{fig:expert-registration}. This metadata is stored in the MCP registry, enabling dynamic invocation at runtime. The expert models are stateless and modular, and can be independently deployed or updated without affecting the LLM.
\item \textbf{LLM-driven expert planning (online)}: Given a user query $q$ and a wireless observation $\mathbf{h}$, the planner prompts the LLM to determine which attributes are relevant. The model selects a subset of experts $\mathcal{E}_\mathcal{A} \subset \{\mathcal{E}_1, \dots, \mathcal{E}_M\}$ for invocation. This decision is based on a semantic understanding of the task and context.

\item \textbf{Expert invocation by MCP client}: For each selected expert $s_m$, the MCP client constructs a JSON-formatted query containing the expert name and the preprocessed input $\mathbf{h} \in \mathbb{R}^n$ (e.g., channel magnitude). The structure of these messages is illustrated in Fig.~\ref{fig:expert-invocation}. The queries are then sent to expert servers, which may be hosted on edge or cloud infrastructure.

\item \textbf{Expert response handling}: Each server returns a standardized response:
\begin{align}
&\texttt{\{confidence: } p(s_m \mid \mathbf{h}), \notag\\
&\texttt{status: OK, source\_id: $m$\}}. \notag
\end{align}
The executor collects and formats the results, which are consistent across all expert endpoints.

\item \textbf{Context augmentation}: The MCP runtime assembles the original query $q$ and all expert outputs and injects them into the LLM's prompt window. This augmentation enables the model to reason using both linguistic instruction and wireless environment information.

\item \textbf{Final reasoning and response generation}: Based on the enriched context, the LLM generates the output $r_\pi(q, \mathbf{h})$. Since the expert outputs are interpretable and structured, the reasoning remains grounded in wireless states without requiring LLM retraining.
\end{enumerate}

This six-stage architecture provides a unified runtime for tool-augmented reasoning in wireless systems. The design supports extensibility, as new experts can be registered without retraining or reconfiguring the LLM. The LLM remains a general-purpose reasoning agent, while the physical interpretation is handled by task-specific, composable expert models orchestrated through the MCP~\cite{anthropic_mcp_2025}. This division of roles enables efficient, interpretable, and domain-adaptive reasoning in dynamic wireless environments.

\begin{figure}[t]
\centering
\begin{tcolorbox}[colback=white!99!gray!1, colframe=black!75, title=Expert Invocation and Response Structure]
{\small \begin{algorithmic}[1]
\State \textbf{function} \textsc{CallExpert}($s_m$, $\mathbf{h}$)
\State \hspace{0.3cm} \textbf{define} query $\gets$
\State \hspace{0.6cm} \{
\State \hspace{1.0cm} \texttt{"tool\_name"}: $s_m$,
\State \hspace{1.0cm} \texttt{"arguments"}: \{
\State \hspace{1.3cm} \texttt{"h"}: vectorized input $\mathbf{h} \in \mathbb{R}^{n}$
\State \hspace{1.0cm} \}
\State \hspace{0.6cm} \}
\State \hspace{0.3cm} Send query to MCP server and await result
\\
\State \hspace{0.3cm} \textbf{receive} response $\gets$
\State \hspace{0.6cm} \{
\State \hspace{1.0cm} \texttt{"confidence"}: $p(s_m \mid \mathbf{h})$,
\State \hspace{1.0cm} \texttt{"status"}: \texttt{"OK"},
\State \hspace{1.0cm} \texttt{"source\_id"}: $m$
\State \hspace{0.6cm} \}
\State \hspace{0.3cm} \textbf{return} response
\State \textbf{end function}
\end{algorithmic}}
\end{tcolorbox}
\caption{Expert invocation procedure. Given a preprocessed input $\mathbf{h}$ and selected tool $s_m$, the MCP client constructs a standardized query. The expert server returns a structured JSON result to be injected into the LLM context.}
\label{fig:expert-invocation}
\end{figure}


\section{Experiment Results}
We evaluate the proposed system through two complementary stages: (1) learning performance of the individual wireless expert classifiers, and (2) end-to-end decision accuracy of the LLM agent with and without MCP integration.

\subsection{Expert Learning Performance}

We first examine four representative experts trained to detect (1) LoS propagation, (2) high Doppler shift, (3) Rayleigh fading, and (4) Rician fading with $K=10$. These classifiers form part of the IoX and are implemented as lightweight MLPs. Each model is trained from synthetic wireless traces constructed to reflect its target condition, using binary cross-entropy loss over $4000$ epochs. Input features are real-valued vectors derived from channel impulse responses.

\begin{figure*}[t]
\centering
\includegraphics[width = 1.0\textwidth]{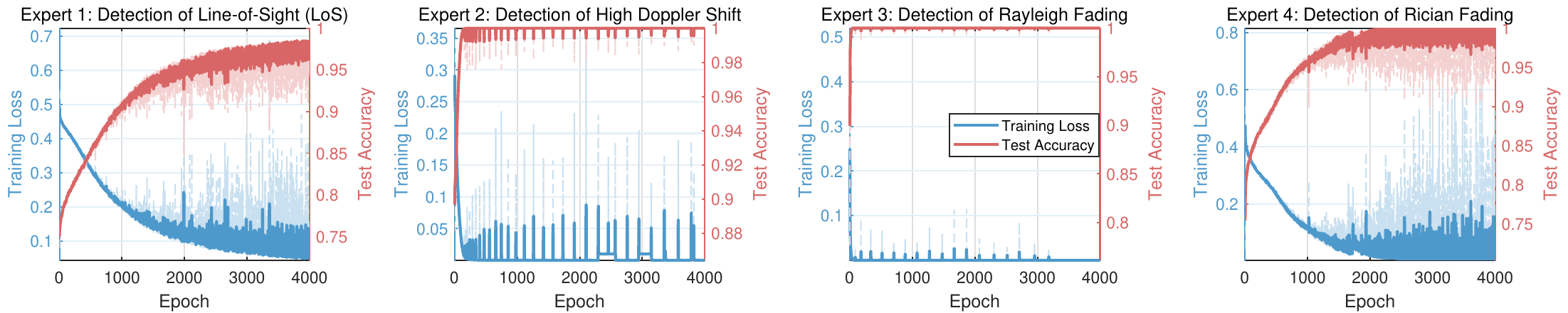}
\vspace{-0.6cm}
\caption{Training and validation performance of three representative expert classifiers. Each subplot shows the evolution of training loss and test accuracy across epochs for detecting (1) LoS conditions, (2) high Doppler shifts, (3) Rayleigh fading, and (4) Rician fading.}
\label{fig_lossacc}
\end{figure*}

Figure~\ref{fig_lossacc} shows the training loss and test accuracy curves for these representative experts. Raw measurements are shown as dashed lines, while smoothed values are plotted as solid curves using a moving average filter. Rayleigh and high Doppler experts converge quickly, reaching high generalization accuracy due to their well-separated physical patterns. The LoS and Rician experts show more gradual improvement, reflecting the greater ambiguity in intermediate fading scenarios. These results confirm that each expert can reliably classify their assigned wireless environment condition and serve as an accurate and stable query endpoint for LLM agents via MCP.

\subsection{End-to-End Agent Performance Evaluation}

To assess the full system, we construct a test set of $1000$ synthetic wireless observations using a randomized channel simulation pipeline. Each sample is annotated with binary labels for several scene attributes. Specifically, the channel responses are generated using a mixture of Rayleigh and Rician components, incorporating time-varying Doppler effects. The input to the LLM is a real-valued vector derived from the magnitude of the complex channel response.

We evaluate two configurations of the LLM agents: a standalone version without tool access and a version enhanced by MCP with expert querying. For each test case, the agent receives a structured prompt:
\begin{quote}
\texttt{"You are a wireless environment reasoning assistant. Given a real-valued channel vector $\mathbf{h}$, infer whether the scene satisfies each of the following attributes: \{\texttt{attribute\_list}\}. Respond only in strict JSON format: \{\texttt{attribute\_json}\}"}
\end{quote}

Here, \texttt{attribute\_list} and \texttt{attribute\_json} are placeholders filled based on the relevant expert set for each task. For example, the placeholders would expand to:
\begin{itemize}
    \item \texttt{attribute\_list}: \texttt{line-of-sight, high Doppler, Rician fading with K = 10}
    \item \texttt{attribute\_json}: \texttt{\{"line-of-sight": 0 or 1, "highdoppler": 0 or 1, "rician\_m10": 0 or 1\}}
\end{itemize}
This templated design supports easy extension to additional attributes by modifying the expert set.

\begin{table}[t]
\caption{Comparison of LLM agents' average accuracy on wireless attribute prediction, with and without MCP.}
\label{tab:llm-acc}
\centering
\setlength{\tabcolsep}{10pt}
\begin{tabular}{l@{\hskip 12pt}c@{\hskip 12pt}c}
\toprule
\textbf{LLM agent} & \textbf{Raw accuracy} & \textbf{+MCP} \\
 & (\%) & (\%) \\
\midrule
\text{DeepSeek-Chat} & 46.7 & 95.5 \\
\text{DeepSeek-Reasoner} & 54.2 & 97.5 \\
\text{ChatGPT-3.5} & 46.6 & 95.8 \\
\text{ChatGPT-4} & 53.5 & 96.9 \\
\text{O4-Mini} & 59.2 & 98.1 \\
\text{QWQ-Plus} & 53.8 & 98.0 \\
\text{Qwen-Plus} & 51.2 & 96.1 \\
\text{Qwen-Turbo} & 45.8 & 95.8 \\
\bottomrule
\end{tabular}
\end{table}

Table~\ref{tab:llm-acc} compares the end-to-end classification accuracy of various LLM agents~\cite{bi2024deepseek}, both in their standalone form and when augmented with MCP. Without MCP, the LLM agents must infer directly from the numerical channel vector $\mathbf{h}$ without any structured assistance. Accuracy in this setting remains modest, typically between $45\%$ and $59\%$. LLMs with enhanced inference-time scaling ability, such as {\textit{DeepSeek-Reasoner}}, {\textit{O4-Mini}}, and {\textit{QWQ}}, outperform their base versions by several percentage points, suggesting that better instruction-following and internal logic improve performance to some extent. However, these gains are still limited, indicating that reasoning via language alone cannot bridge the representational gap between raw physical-layer wireless environment inputs and deterministic tasks. 
In contrast, when MCP is enabled, accuracy jumps significantly across all LLMs, each surpassing $95\%$. This improvement is attributable to the integration of IoX-generated confidence scores, which provide the LLM with structured, high-precision interpretations of wireless conditions. 
As a result, even LLM agents with weaker standalone performance, such as {\textit{Qwen-Turbo}} or {\textit{ChatGPT-3.5}}, match or exceed the best raw-inference agents when supported by expert outputs. Despite minor errors from MCP call formatting inconsistencies or borderline expert predictions, the overall performance approaches the classification accuracy ceiling determined by the experts themselves. 

\section{Conclusion}
We proposed an MCP-based IoX framework to equip LLM agents with wireless environment awareness. By decoupling high-level reasoning from physical-layer environment interpretation, our design enables LLMs to selectively invoke lightweight expert classifiers at inference time, without retraining or embedding domain-specific priors. We formalized the system design, implemented modular expert models for key wireless attributes, and developed an MCP runtime for structured expert querying. Experimental results across several LLMs show that the proposed architecture achieves significant performance gains, improving classification accuracy from $45\%$–$59\%$ to over $95\%$, highlighting the benefit of structured wireless perception in LLM-based agents. 

The framework supports modularity, extensibility, and real-time operation, offering a practical direction for integrating reasoning agents into dynamic wireless networks. 
Future work will evaluate the framework on over‑the‑air channel measurements, profile end‑to‑end latency under strict scheduling budgets, and study cost‑aware expert selection and security safeguards for large‑scale deployment.

\bibliographystyle{IEEEtran}
\bibliography{Ref}

\begin{thebibliography}{10}
\providecommand{\url}[1]{#1}
\csname url@samestyle\endcsname
\providecommand{\newblock}{\relax}
\providecommand{\bibinfo}[2]{#2}
\providecommand{\BIBentrySTDinterwordspacing}{\spaceskip=0pt\relax}
\providecommand{\BIBentryALTinterwordstretchfactor}{4}
\providecommand{\BIBentryALTinterwordspacing}{\spaceskip=\fontdimen2\font plus
\BIBentryALTinterwordstretchfactor\fontdimen3\font minus \fontdimen4\font\relax}
\providecommand{\BIBforeignlanguage}[2]{{%
\expandafter\ifx\csname l@#1\endcsname\relax
\typeout{** WARNING: IEEEtran.bst: No hyphenation pattern has been}%
\typeout{** loaded for the language `#1'. Using the pattern for}%
\typeout{** the default language instead.}%
\else
\language=\csname l@#1\endcsname
\fi
#2}}
\providecommand{\BIBdecl}{\relax}
\BIBdecl

\bibitem{zhou2024large}
H.~Zhou, C.~Hu, Y.~Yuan, Y.~Cui, Y.~Jin, C.~Chen, H.~Wu, D.~Yuan, L.~Jiang, D.~Wu \emph{et~al.}, ``Large language model ({LLM}) for telecommunications: A comprehensive survey on principles, key techniques, and opportunities,'' \emph{IEEE Commun. Surv. Tutor.}, to appear, 2025.

\bibitem{bi2024deepseek}
X.~Bi, D.~Chen, G.~Chen, S.~Chen, D.~Dai, C.~Deng, H.~Ding, K.~Dong, Q.~Du, Z.~Fu \emph{et~al.}, ``Deepseek {LLM}: Scaling open-source language models with longtermism,'' \emph{{\rm arXiv preprint arXiv:2401.02954}}, 2024.

\bibitem{wang2024survey}
L.~Wang, C.~Ma, X.~Feng, Z.~Zhang, H.~Yang, J.~Zhang, Z.~Chen, J.~Tang, X.~Chen, Y.~Lin \emph{et~al.}, ``A survey on large language model-based autonomous agents,'' \emph{Front. Comput. Sci.}, vol.~18, no.~6, p. 186345, June 2024.

\bibitem{zhou2025large}
H.~Zhou, C.~Hu, D.~Yuan, Y.~Yuan, D.~Wu, X.~Chen, H.~Tabassum, and X.~Liu, ``Large language models for wireless networks: An overview from the prompt engineering perspective,'' \emph{IEEE Wireless Commun.}, 2025.

\bibitem{liu2025llm4wm}
X.~Liu, S.~Gao, B.~Liu, X.~Cheng, and L.~Yang, ``{LLM4WM}: Adapting {LLM} for wireless multi-tasking,'' \emph{{\rm arXiv preprint arXiv:2501.12983}}, 2025.

\bibitem{noh2025adaptive}
H.~Noh, B.~Shim, and H.~J. Yang, ``Adaptive resource allocation optimization using large language models in dynamic wireless environments,'' \emph{{\rm{arXiv preprint arXiv:2502.02287}}}, 2025.

\bibitem{du2024reinforcement}
H.~Du, R.~Zhang, D.~Niyato, J.~Kang, Z.~Xiong, and D.~I. Kim, ``Reinforcement learning with large language models ({LLMs}) interaction for network services,'' in \emph{Proc. International Conference on Computing, Networking and Communications (ICNC)}.\hskip 1em plus 0.5em minus 0.4em\relax IEEE, 2024, pp. 799--803.

\bibitem{hou2025model}
X.~Hou, Y.~Zhao, S.~Wang, and H.~Wang, ``Model context protocol ({MCP}): Landscape, security threats, and future research directions,'' \emph{{\rm{arXiv preprint arXiv:2503.23278}}}, 2025.

\bibitem{anthropic_mcp_2025}
{Anthropic}, ``Model context protocol,'' \url{https://www.anthropic.com/news/model-context-protocol}, 2025, accessed: 27~Apr.~2025.

\bibitem{barati2016initial}
C.~N. Barati, S.~A. Hosseini, M.~Mezzavilla, T.~Korakis, S.~S. Panwar, S.~Rangan, and M.~Zorzi, ``Initial access in millimeter wave cellular systems,'' \emph{IEEE Trans. Wireless Commun.}, vol.~15, no.~12, pp. 7926--7940, Dec. 2016.

\bibitem{zappone2019model}
A.~Zappone, M.~Di~Renzo, M.~Debbah, T.~T. Lam, and X.~Qian, ``Model-aided wireless artificial intelligence: Embedding expert knowledge in deep neural networks for wireless system optimization,'' \emph{IEEE Veh. Technol. Mag.}, vol.~14, no.~3, pp. 60--69, Mar. 2019.

\bibitem{panic2013fading}
S.~Panic, M.~Stefanovic, J.~Anastasov, and P.~Spalevic, \emph{Fading and interference mitigation in wireless communications}.\hskip 1em plus 0.5em minus 0.4em\relax CRC press, 2013.

\bibitem{ruby2020binary}
U.~Ruby, V.~Yendapalli \emph{et~al.}, ``Binary cross entropy with deep learning technique for image classification,'' \emph{Int. J. Adv. Trends Comput. Sci. Eng}, vol.~9, no.~10, 2020.

\bibitem{qintoolllm}
Y.~Qin, S.~Liang, Y.~Ye, K.~Zhu, L.~Yan, Y.~Lu, Y.~Lin, X.~Cong, X.~Tang, B.~Qian \emph{et~al.}, ``{ToolLLM}: Facilitating large language models to master 16000+ real-world {APIs},'' in \emph{Proc. Int. Conf. Learn. Represent.}, 2024.

\end{thebibliography}
\end{document}